\documentclass{ws-ijmpd}
\usepackage[super,compress]{cite}
\usepackage{color}

\usepackage{multirow}

\usepackage{pifont}
\newcommand{\cmark}{\ding{51}}%
\newcommand{\xmark}{\ding{55}}%

\makeatletter
\newcommand*{\rom}[1]{\expandafter\@slowromancap\romannumeral #1@}
\makeatother

\begin{document}

%
\catchline{}{}{}{}{}
%

\title{The little sibling of the big rip singularity}

\author{Mariam Bouhmadi-L\'{o}pez $^{1,2,3,4\: a}$, Ahmed Errahmani $^{5,\: b}$, Prado Mart\'in-Moruno $^{6,7,\: c}$, Taoufik Ouali $^{5,\: d}$ and Yaser Tavakoli $^{8,\: e}$}

\address{
$^{1}$Departamento de F\'{i}sica,  Universidade da Beira Interior, 6200 Covilh\~a, Portugal\\
$^{2}$Centro de Matem\'atica e Aplica\c{c}\~oes da Universidade da Beira Interior (CMA-UBI), 6200 Covilh\~a, Portugal\\
$^{3}$Department of Theoretical Physics, University of the Basque Country UPV/EHU, P.O. Box 644, 48080 Bilbao, Spain\\
$^{4}$IKERBASQUE, Basque Foundation for Science, 48011, Bilbao, Spain\\
$^{5}$Laboratory of Physics of Matter and Radiation, Mohammed I University, BP 717, Oujda, Morocco\\
$^{6}$Centro de Astronomia e Astrof\'{\i}sica da Universidade de Lisboa, Campo Grande, Edific\'{\i}o C8, 1749-016 Lisboa, Portugal\\
$^{7}$Centro Multidisciplinar de Astrof\'{\i}sica - CENTRA, Departamento de F\'{\i}sica, Instituto Superior T\'ecnico, Av. Rovisco Pais 1,1049-001 Lisboa, Portugal\\
$^{8}$Departamento de F\'isica, Universidade Federal do Esp\'irito Santo, Av. Fernando Ferrari 514, Vit\'oria - ES, Brazil\\
$^{a}$mbl@ubi.pt On leave of absence from UPV and IKERBASQUE\\
$^{b}$ahmederrahmani1@yahoo.fr\\
$^{c}$prado.moruno@tecnico.ulisboa.pt\\
$^{d}$ouali\_ta@yahoo.fr\\
$^{e}$tavakoli@cosmo-ufes.br}


\maketitle

\begin{history}
\received{Day Month Year}
\revised{Day Month Year}
\end{history}

\begin{abstract}

We present a new cosmological event, which we named the little sibling of the big
rip. This event is much smoother than the big rip singularity. When the little sibling
of the big rip is reached, the Hubble rate and the scale factor blow up but the cosmic
derivative of the Hubble rate does not. This abrupt event takes place at an infinite
cosmic time where the scalar curvature explodes. We show that a doomsday  \`{a} la
little sibling of the big rip is compatible with an accelerating universe, indeed at present
it would mimic perfectly a $\Lambda$CDM scenario.
It turns out that even though the event seems to be harmless as it takes place in the infinite future,
the bound structures in the universe would be unavoidably destroyed on a finite cosmic time from now.
The model can be motivated by considering that the  weak energy condition should not be abusively violated in our Universe, and it could give us some hints
about the status of recently formulated non-linear energy conditions.

\end{abstract}


\section{Introduction}

The issue of singularities is a fascinating area of general relativity \cite%
{gravitation,largescale}. The theory is supposed to cease to be valid close
to the singularities and it is expected that a semi-classical approach to
general relativity or a quantum theory of gravity can cure those
singularities. On a cosmological setting the most famous singularities are
the big bang in the past and the big crunch in the future \cite{gravitation}%
. It turns out that the astonishing discovery of the current acceleration of
the universe, as implied by observations of supernovae of type Ia \cite%
{SNeIa} and confirmed by data from the cosmic microwave background \cite{CMB,Planck}
and the baryon acoustic oscillations \cite{BAO} among others, has opened up
the possibility of different future doomsdays of the universe.

Depending on the nature of dark energy, causing the present acceleration of
the universe, and its future behaviour, the universe might face completely
different destinies: given the present data, an asymptotically de
Sitter-like behaviour where the universe is eternally expanding is the
favoured option but not the unique one. If the equation of state of dark
energy, i.e., the ratio between its pressure and energy density, is less
than -1 then dark energy violates effectively the null energy condition and
the universe could face a big rip singularity \cite%
{Caldwell:1999ew,Starobinsky:1999yw,Carroll:2003st,Caldwell:2003vq,Chimento:2003qy,Chimento:2003qy,Dabrowski:2003jm,GonzalezDiaz:2003rf,GonzalezDiaz:2004vq,Nojiri:2004pf}%
, where the scale factor of the universe, the Hubble rate and its cosmic
time derivative blow up in a finite future cosmic time. It was soon realised
that this is not the unique possible doomsday of a dark energy dominated
universe. Indeed, the universe might end up in a sudden singularity \cite%
{Barrow:2004xh,Gorini:2003wa,Nojiri:2005sx} (or even a type IV singularity
\cite{Nojiri:2005sx} if gravity deviates from general relativity at
late-time), a big freeze doomsday \cite%
{Nojiri:2004pf,Nojiri:2005sx,Nojiri:2005sr,BouhmadiLopez:2006fu,BouhmadiLopez:2007qb}
or what has been recently named a little rip \cite{Frampton:2011sp}, even
though previously discovered in general relativity \cite%
{Nojiri:2005sx,Stefancic:2004kb} and modified theories of gravity of the
kind of brane-world \cite{BouhmadiLopez:2005gk}.\\
The purpose of this paper is to present a new event which we named the
little sibling of the big rip, where the scalar curvature blows up due to a
divergence of the Hubble parameter, even though the derivative of the Hubble
parameter is finite at the event. In addition, the little sibling of the big
rip is reached in an infinite cosmic time. We used: (i) the name sibling
because like on a big rip singularity, the Hubble rate diverges and (ii)
little because it takes the universe an infinite cosmic time to reach the
event like what happens with the little rip event. We will show as well how
this model incorporates in a natural way an accelerating phase for a
homogeneous and isotropic universe.\\
The paper is outlined as follows. In the next section, we introduce the model and discuss its cosmological solutions.
In section III,  we present a clear motivation of the model from the point of view of
some energy conditions, and study the behaviour of the model regarding other energy conditions.
Then, in section IV, we discuss the outcome of bound structure when the universe approaches the  little sibling of the big rip
and finally we conclude and summarise our results in section V.

\section{The little sibling of the big rip}
\label{secII}

We start considering a spatially flat Friedmann-Lema\^{\i}%
tre-Robertson-Walker (FLRW) universe filled with (dark) matter and dark energy with energy densities  $\rho_{\rm matt}$ and $\rho_{\rm DE}$, respectively%
\footnote{%
For simplicity, we have disregarded the radiation component as we are mainly
interested on the late-time acceleration of the universe and its asymptotic
future behaviour. None of the conclusions presented in this paper are
modified by the inclusion of the radiation component.}.
The evolution equations are given by the Friedmann equation
\begin{equation}
H^{2}=\frac{8\pi G}{3}\left(\rho_{\rm matt}+\rho_{\rm DE}\right) ,
\end{equation}%
and the Raychaudhuri equation
\begin{align}
\frac{\ddot{a}}{a}\  =\  -\frac{4\pi G}{3}\left[\rho_{\rm matt} + \rho_{\rm DE} +3p_{\rm DE} \right]\ ,
\label{Raychadhuri}
\end{align}
where $G$ is the gravitational constant.  We
assume a dark energy component whose equation of state deviates slightly
from a cosmological constant as follows\footnote{%
A more general equation of state was considered in Ref.~\refcite{Alcaniz:2005tw}
which includes as a particular case Eq.~(\ref{eqstate}). Here we perform a full analysis
of the cosmological background evolution of the model, studying the fate of bound structures
in this model and giving a clear motivation for analysing this model form the point of view of
the energy conditions.}
\begin{equation}
p_{\rm DE}+\rho _{\rm DE}\ =\ -\frac{A}{3}\ ,  \label{eqstate}
\end{equation}%
where $A$ is a positive constant and the coefficient $1/3$ has been
introduced for latter convenience. The energy momentum tensor of the perfect
fluid describing dark energy is conserved, implying therefore that
$\rho_{\rm DE}$ evolves with the scale factor, $a$, as
\begin{equation}
\rho _{\rm DE}\ =\ \Lambda +A\ln \left( \frac{a}{a_{0}}\right) ,  \label{desibling}
\end{equation}%
being $\Lambda $ an integration constant, playing the role of an effective
cosmological constant at present and which we will assume positive, and $%
a_{0}$ standing for the present scale factor of the universe. From now on,
the subscript $0$ stands for quantities evaluated at the present time.

The relation (\ref{desibling}) shows that the energy density $\rho_{\rm DE}$
describes a constant energy density with a logarithmic correction of the
scale factor. As we will show a tiny deviation of the equation of state (\ref%
{eqstate}) from that of a cosmological constant; i.e. a tiny constant $A$ in
Eq.~(\ref{eqstate}), implies a little sibling of the big rip on the future
evolution of the universe. More precisely, we will show that the Hubble rate
and the scale factor blow up in an infinite cosmic time while the cosmic
derivative of the Hubble rate remains finite.

\begin{figure}[h]
\begin{center}
\includegraphics[width=0.55\columnwidth]{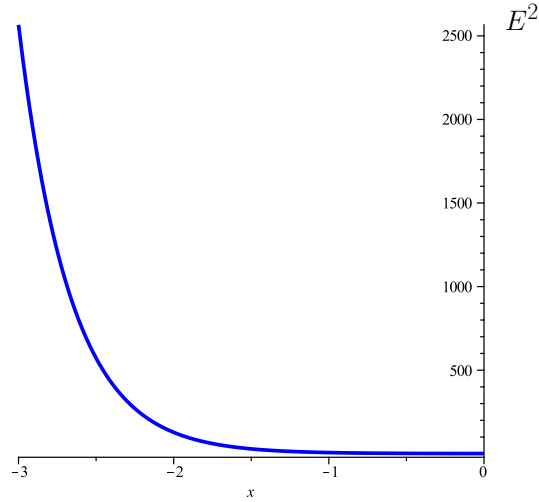}
\end{center}
\caption{The blue solid line represents the square of the dimensionless
Hubble parameter; i.e. $E^2$ given in Eq.~(\protect\ref{friedmannx}), as a
function of the variable $x$ related to the scale factor through $x=\ln(a/a_0)$. The plot has been done setting $\Omega_m=0.315$,
corresponding to the value obtained by the Planck collaboration \cite{Planck}, and $\Omega_x=10^{-3}$.}
\label{phubble}
\end{figure}
The Friedmann and  the Raychaudhuri equations can be written as
\begin{align}
E^2\ & =\  \Omega_m e^{-3x} +\Omega_\Lambda + \Omega_x x,  \label{friedmannx} \\
\frac{\ddot{a}}{a}\  &=\  -\frac{H_0^2}{2}\left[\Omega_me^{-3x}-2\Omega_\Lambda-\Omega_x(1+2x)\right]\ ,
\label{Raychadhuri-b}
\end{align}
where $E=H/H_0$ is the dimensionless Hubble rate, and
\begin{align}
& \Omega_m := \frac{\rho_{\rm matt}^0}{3M_{\rm Pl}^2H_0^2}\ , \  \  \  \  \  \  \  \   \Omega_\Lambda :=  \frac{\rho_{\Lambda}}{3M_{\rm Pl}^2H_0^2}\ ,  \notag \\
& \Omega_x :=  \frac{A}{3M_{\rm Pl}^2H_0^2}\ ,  \  \  \  \  \  \  \  \  \ x := \ln \left(\frac{a}{a_0}\right)\ ,
\end{align}
with  $8\pi G=M_{\rm Pl}^{-2}$ being the reduced Planck mass.
In Fig. \ref{phubble} we show the evolution of $E^2$ for a particular model.
Evaluating the Friedmann equation (\ref{friedmannx}) at present; i.e. $x=0$,
we obtain the constraint
\begin{equation}
\Omega_m+\Omega_\Lambda=1\  .
\end{equation}
This model mimics the $\Lambda$CDM scenario at present as long as $%
|\Omega_x|\ll 1$ which we will assume it is the case. Indeed, as long as $%
\Omega_x\ll 1$ this model mimics a flat $\Lambda$CDM model till now, even
though its future behaviour might be quite different. Please notice as well
that although the dark energy component (\ref{desibling}) becomes negative
in the past, it is always sub-dominant with respect to dark matter until the
future, at least for physically reasonable values of $\Omega_x$ such that
the model does not deviate too much from the $\Lambda$CDM (we have assumed $%
\Omega_x\lesssim 10^{-1}$). Notice that a similar reasoning applies to the
radiation component of the universe. In fact, for the range of values
assumed for $\Omega_x$, the model is at present pretty much similar to the $%
\Lambda$CDM scenario while the dark energy component is sub-dominant at the
big bang nucleosynthesis period and at the matter domination period.

At present and in the recent past the modified Friedmann equation can be
approximated by
\begin{equation}
E^{2}\  \approx\  \Omega _{m}e^{-3x}+\Omega _{\Lambda }\ ,  \label{friedmannpast}
\end{equation}%
whose solution can be expressed as
\begin{align}
a(t) \  \approx\   a_{0} \left( \frac{\Omega _{m}}{\Omega _{\Lambda }}%
\right) ^{\frac{1}{3}}  \sinh ^{\frac{2}{3}}\left[\frac{3}{2}H_{0}\sqrt{\Omega _{\Lambda }}%
(t-t_{0})-\frac{1}{2}\ln \left( \frac{1-\sqrt{\Omega _{\Lambda }}}{1+\sqrt{%
\Omega _{\Lambda }}}\right) \right].
 \label{Solfriedmannpast}
\end{align}

On the future, the asymptotic behaviour of the Friedmann equation can be
simplified as
\begin{equation}
E^2\  \approx\  \Omega_\Lambda + \Omega_x x,  \label{friedmannfuture}
\end{equation}
whose solution is obtained as
\begin{align}
a(t)   \  \approx \    a_0e^{-\frac{\Omega_\Lambda}{\Omega_x}}  \exp\left\{\frac{1}{\Omega_x}\left[\frac{\Omega_x}{2}H_0(t-t_1)+\sqrt{\Omega_xx_1+\Omega_\Lambda}\right]^2\right\},
\label{scalefactor2}
\end{align}
where $t_1$, $x_1=\ln(a_1/a_0)$ and $a_1$ are constants. Therefore, we can
rewrite the dimensionless Hubble parameter; E, as
\begin{equation}
E\  \approx\  \frac{\Omega_x}{2}H_0(t-t_1) +\sqrt{\Omega_\Lambda+\Omega_x x_1}\ .
\label{Hubbleinfinity}
\end{equation}
Consequently, the Hubble rate blows up when the cosmic time gets very large.
Amazingly, the time derivative of the Hubble rate remains finite as can be
easily shown from Eq.~(\ref{Hubbleinfinity}), more explicitly
\begin{equation}
\dot{E}\  \approx\  \frac{\Omega_x}{2}H_0\ .
\end{equation}
This event is precisely what we named the little sibling of the big rip. It
is an event where the Hubble rate blows up as well as the scale factor,
while the cosmic derivative of the Hubble rate is finite. Consequently, the
scalar curvature blows up at this event which takes place in an infinite
cosmic time. We can show as well that the equation of state of dark energy, $%
w_{\rm DE}$, approaches $-1$. Indeed using Eq.~(\ref{eqstate}), we obtain
\begin{equation}\label{w}
w_{\rm DE}\ =\ -1-\frac{A}{3\rho_{\rm DE}}\ .
\end{equation}
As the energy density, $\rho_{\rm DE}$, gets infinitely large in the future, the
equation of state parameter $w_{\rm DE}$ behaves as the one of a cosmological
constant. Consequently, the little sibling of the big rip is much smoother
than the little rip and even more than a big rip. Please notice that even though the equation of state $w_{\rm DE}$ approaches -1 at infinity,
the universe is not asymptotically de Sitter because $\rho_{\rm DE}$ blows up
as can be easily noticed by considering the conservation equation for the fluid (\ref{eqstate}).
An alternative way of proving that the asymptotic behaviour does not correspond to a de Sitter universe is by obtaining the asymptotic value of the
scalar curvature on that regime, $R\approx 8\pi G (A+4\rho)$ which clearly blows up for large $\rho$, as it is precisely the case.

\begin{figure}[t!]
\begin{center}
\includegraphics[width=0.5\columnwidth]{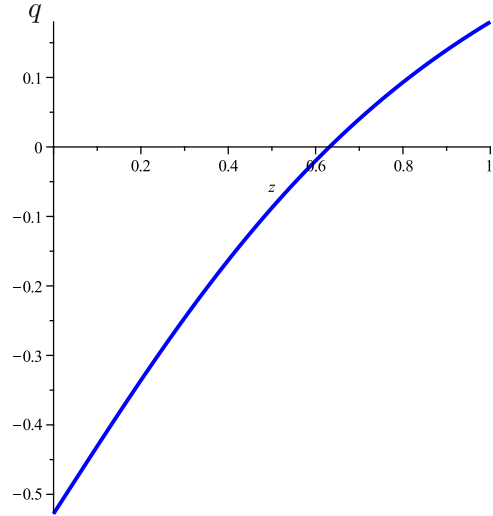}
\end{center}
\caption{The blue solid line represents the square of the deceleration
parameter; i.e. $q=-(1+\dot{H}/H^2)$ given in Eq.~(\protect\ref{friedmannx}%
), as a function of the variable $z$ related to $x$ as $z=1-\exp(-x)$. The
plot has been done setting $\Omega_m=0.315$, corresponding to its Planck
value \protect\cite{Planck}, and $\Omega_x=10^{-3}$.}
\label{pq}
\end{figure}

Can the dark energy model introduced in Eq.~(\ref{desibling}) describe the
present acceleration of the universe? The deceleration parameter reads
\begin{equation}
q\ :=\ -\left( 1+\frac{1}{2E^{2}}\frac{dE^{2}}{dx}\right)\ ,
\label{defdecceleration}
\end{equation}%
and in particular at present reduces to 
\begin{equation}
q_{0}\ =\ -\left(1- \frac{3}{2}\Omega _{m}+\frac{1}{2}\Omega _{x}\right)\  .
\label{defdecceleration}
\end{equation}
Therefore, the universe is at present accelerating as long as $%
3\Omega_{m}-\Omega _{x}<2$, condition which is fulfilled for the present
amount of dark energy and matter $\Omega _{m}\simeq 0.315$ {\cite{Planck}}. In
addition, the deviation of the dark energy model Eq.~(\ref{desibling}) from
the $\Lambda $CDM, quantified on $\Omega _{x}$, favours the current
acceleration of the universe as compared with the pure $\Lambda $CDM model
as shown in Eq.~(\ref{defdecceleration}). As an example, we show on Fig.~\ref%
{pq} the evolution of the deceleration parameter $q$ as a
function of the redshift.

\begin{figure}[h]
\begin{center}
\includegraphics[width=0.55\columnwidth]{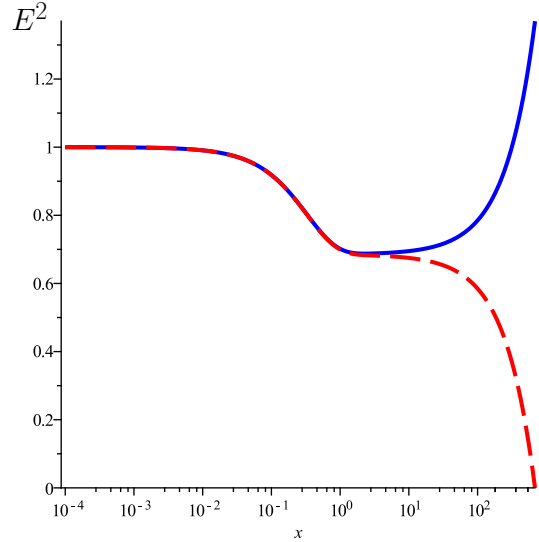}
\end{center}
\caption{The blue solid line and red dashed line represent the square of the dimensionless
Hubble parameter; i.e. $E^2$ given in Eq.~(\protect\ref{friedmannx}), as a
function of the variable $x$ related to the scale factor through $%
x=\ln(a/a_0)$ for $\Omega_x=10^{-3}$ and $\Omega_x=-10^{-3}$, respectively.
The plot has been done setting $\Omega_m=0.315$, corresponding to its Planck
value \protect\cite{Planck}. For negative values of $\Omega_x$
the universe hits a bounce while for positive $\Omega_x$ the universe
expands for ever ending in a little sibling of the big rip.}
\label{phubble2}
\end{figure}

For completeness, we analyse what happens for a similar model of dark energy
as the one described in Eq.~(\ref{eqstate}) but with a negative $A$ or
equivalently $\Omega _{x}<0$. The past expansion of the universe is
unaffected by the sign of $A$, as long as its absolute value is within the
range mentioned above. However, the future
evolution of the universe depends strongly on the sign of $A$ (cf. Fig.~\ref{phubble2}). Indeed, if
the dark energy density (\ref{eqstate}) fulfils the null energy condition ($%
A<0 $), the universe bounces at some point in the future where {the scale
factor can be approximated as $\ln (a/a_{0})\simeq -\Omega _{\Lambda
}/\Omega _{x}$}, however if the dark energy (\ref{eqstate}) does not fulfil
the null energy condition, the universe hits a little sibling of the big rip
singularity in its future. 
In fact in the case $A<0,$ the solution of Eq.~(\ref{friedmannfuture}) is
given by (\ref{scalefactor2}) with $\Omega _{x}<0$, and $x_{1}<-\Omega
_{\Lambda }/\Omega _{x}$. The universe bounces at $x_{b}\simeq -\Omega
_{\Lambda }/\Omega _{x}=-\Lambda /A$, when
\begin{equation}
t_{b}\ \simeq\   t_{1}-\frac{2}{\Omega _{x}H_{0}}\sqrt{\ \Omega
_{x}x_{1}+\Omega _{\Lambda }}\ .
\end{equation}%
Plugging the previous relation in Eq (\ref{scalefactor2}), we can rewrite the
log of the scale factor as
\begin{equation}
x\  \simeq\  \frac{\Omega _{x}}{4}H_{0}^{2}(t-t_{b})^{2}-\frac{\Omega _{\Lambda
}}{\Omega _{x}}\ ,
\end{equation}%
which results in the following dimensionless Hubble parameter
\begin{equation}
E\ \simeq \ \frac{\Omega _{x}}{2}H_{0}(t-t_{b})\ .
\end{equation}%

The previous solution shows that the universe would contract just after the bounce {$%
(t_{b}<t)$ returning eventually to its \textquotedblleft
initial\textquotedblright\ state when $a\rightarrow 0$}. This event can be
explained by the fact that the dark energy (\ref{desibling}), which is
responsible for the current accelerated expansion of the universe,
decreases with the expansion until it reaches a vanishing value when $%
x_{b}\simeq -\Lambda /A$ and where $\rho_m$ can be neglected. At this time the universe bounce starting its
contraction, this contraction becomes more and more violent.
Indeed the deceleration parameter (\ref{defdecceleration}) which is
currently roughly of the order $-0.528\,$ can be approximated around the bounce
as:
\begin{equation}
q\ =\ -\left( 1+\frac{2}{\Omega _{x}H_{0}^{2}(t-t_{b})^{2}}\right)\ .
\end{equation}%
Consequently, the deceleration parameter increases with the accelerated
expansion and becomes less negative in the future until vanishing at $t_{1}^{\prime }=t_{b}-$ $\frac{1}{H_{0}}\sqrt{\frac{2}{%
|\Omega _{x}|}}$, which correspond to $x\simeq x_{b}-1/2$, and where $q$ changes its sign. Afterwards $%
(t>t_{1}^{\prime })$ the universe enters a phase of decelerated expansion
and the deceleration parameter gets increasingly positive until it
diverges at the bounce, i.e. at $t=t_{b}$ or at $x=x_{b}$. For $%
t>t_{b}$, the deceleration parameter changes its monotony and decreases to
less positive values until it vanishes and changes its sign once more time
for $t_{2}^{\prime }=t_{b}+$ $\frac{1}{H_{0}}\sqrt{\frac{2}{|\Omega _{x}|}}$.
In this range the contraction of the universe is decelerated. For $%
t>t_{2}^{\prime }$ the deceleration parameter becomes more and more negative
converging to $-1$ which correspond to an accelerated contraction. This
accelerated contraction is promoted by the exponential growth of the density of matter
which becomes increasingly important, in this situation the asymptotic
behaviour of the Friedmann equation (\ref{friedmannfuture}) is no longer
valid. Consequently, the competition on the dynamics of the universe between
dark energy and cold dark matter leads to a situation similar to our recent past or present where the
modified Friedmann equation can be  approximated by Eq.~(\ref{friedmannpast}) which
gives the solution (\ref{Solfriedmannpast}) but in this case in a contracting way.

\section{Energy conditions}

The consideration of a dark fluid; i.e. a dark energy component, given by equation (\ref{eqstate}) has a clear motivation coming from the
recently formulated quantum version of the energy conditions \cite{Martin-Moruno:2013wfa}, although its
behaviour in relation to these conditions is not at all trivial as we will show in this section.

Let us start summarising the linear point-like energy conditions.
As is well known these conditions are not fundamental physics \cite{Barcelo:2002bv};
however, they are useful to characterise the kind of fluid
we are dealing with. These conditions are \cite{largescale,Matt}:
\begin{itemize}
 \item Strong energy condition (SEC): Gravity should always be attractive.
In a cosmological scenario implies
\begin{equation}\label{SEC}
 {\rm SEC:}\qquad\rho+3p\geq0.
\end{equation}

 \item Dominant energy condition (DEC): The energy density measured by any observer
should be non-negative and propagate in a causal
way, which leads to
\begin{equation}\label{DEC}
 {\rm DEC:}\qquad\rho\geq |p|.
\end{equation}

\item Weak energy condition (WEC): The energy density measured by any observer should be non-negative
\begin{equation}\label{WEC}
 {\rm WEC:}\qquad\rho\geq 0, \qquad {\rm and}\qquad \rho+p\geq0.
\end{equation}

\item Null energy condition (NEC): A minimum requirement which is implied by SEC and WEC, is
\begin{equation}\label{NEC}
 {\rm NEC:}\qquad \rho+p\geq0.
\end{equation}
Therefore, its violation implies that none of the mentioned energy conditions are satisfied.
\end{itemize}
The SEC is known to be violated in classical scenarios  \cite{Barcelo:2002bv} and in particular it has to be
violated in an accelerating universe like the primordial inflationary phase or the current one. Moreover, violations of all
these conditions are known when considering quantum effects  \cite{Barcelo:2002bv}. Thus,
non-linear constraints on the stress-energy tensor have been introduced recently
in the literature \cite{Abreu:2011fr,Martin-Moruno:2013sfa,Martin-Moruno:2013wfa}. It has been shown that
they behave better in the presence of semi-classical
quantum effects, although the potential physical interpretation of some of these conditions is not so clear as those
of their linear counterparts \cite{Martin-Moruno:2013wfa}. These non-linear conditions are
\cite{Abreu:2011fr,Martin-Moruno:2013sfa,Martin-Moruno:2013wfa}:
\begin{itemize}
 \item Flux energy condition (FEC): The energy density measured by any observer
should propagate in a causal way (without imposing any restriction on its value).
This leads to \cite{Abreu:2011fr,Martin-Moruno:2013sfa}
\begin{equation}\label{FEC}
 {\rm FEC:}\qquad\rho^2\geq p^2,
\end{equation}
in a cosmological scenario.
\item Determinant energy condition (DETEC): The determinant of the stress-energy tensor is non-negative, then
\cite{Martin-Moruno:2013wfa}
\begin{equation}\label{DETEC}
 {\rm DETEC:}\qquad \rho p^3\geq0.
\end{equation}

\item Trace-of-square energy condition (TOSEC):
The trace of the squared stress-energy tensor is non-negative, which implies \cite{Martin-Moruno:2013wfa}
\begin{equation}\label{TOSEC}
 {\rm TOSEC:}\qquad \rho^2+3\,p^2\geq0.
\end{equation}
Thus, it is always satisfied in cosmological scenarios.
\end{itemize}

Moreover, the quantum version of these non-linear energy conditions seem to be
satisfied even for quantum vacuum states. Due to its interesting interpretation the
quantum version of the FEC is of particular relevance. This is \cite{Martin-Moruno:2013sfa}
\begin{itemize}
 \item Quantum flux energy condition (QFEC): the energy should either propagate in a causal
way or have associated a flux vector with a norm bounded from above by a quantity depending on the characteristics
of the system under consideration (a characteristics distance of
the system $L$ and  the system 4-velocity $U^a$). In a cosmological scenario it would be natural to consider\footnote{Other distances can be considered as happens for example on the holographic dark energy scenario (see for example \refcite{Ouali:2011,Ouali:2012,BouhmadiLopez:2013pn}.).}
$L=H_0^{-1}$; thus, this condition is expressed as \cite{Martin-Moruno:2013wfa}
\begin{equation}\label{QFEC}
 {\rm QFEC:}\qquad\rho^2-p^2\geq-\epsilon^2,
\end{equation}
with
\begin{equation}\label{epsilon}
\epsilon=\hbar H_0^4\sim 10^{-112}J\,s^{-3}.
\end{equation}
\end{itemize}
As it has been pointed out in Ref.~\refcite{Martin-Moruno:2013wfa}, a similar quantum formulation of the linear energy
conditions would not be so widely satisfied when considering semi-classical quantum effects (although they
could be fulfilled in some situations where their classical counterparts are violated,
for example, for the Casimir vacuum \cite{Martin-Moruno:2013wfa}).
A quantum version
of the WEC has been specifically formulated as \cite{Martin-Moruno:2013wfa}
\begin{itemize}
\item Quantum weak energy condition (QWEC): The energy density measured by any observer should not be
excessively negative, which in a cosmological scenario would lead to
\begin{equation}\label{QWEC}
 {\rm QWEC:}\qquad\rho\geq -\epsilon, \qquad {\rm and}\qquad \rho+p\geq-\epsilon.
\end{equation}
\end{itemize}
As the total energy density of the model presented in the previous section,
$\rho_{\rm m}+\rho_{\rm DE}$, should always be positive
to have a well defined Hubble parameter,
the total content of the universe we are considering could satisfy the QWEC for
small enough values of the parameter $A$, as one may have
\begin{equation}
 \rho_{\rm m}+\rho_{\rm DE}+p_{\rm DE}\geq-A/3\geq-\epsilon.
\end{equation}
Moreover, one could use these quantum conditions to quantify how much the classical energy
conditions are violated.
Therefore, the dark energy of this paper could be interpreted as
a fluid which violates the NEC ``minimally'' if $A$ is small enough.
It must be noted that a covariant formulation of
the quantum NEC in the way of the QWEC introduced in Ref.~\refcite{Martin-Moruno:2013wfa} is not possible
(due to an arbitrariness
in a normalisation factor); however, once the NEC is formulated for a particular stress-energy tensor (as it
is done in (\ref{NEC})) a natural idea of what is a minimal violation could be considered taking the inequality
on the right
in (\ref{QWEC}).

\subsection{Phantom fluid $A>0$}
From the equation of state (\ref{eqstate}), it is clear that this fluid violates the WEC and the NEC.
As it has already been mentioned the dark fluid of this cosmological model
has a negative energy density in the past history of the universe,
where it is a sub-dominant component.
Taking into account (\ref{w}), it can be concluded that the SEC (\ref{SEC})
associated with the dark fluid is violated when
$\rho_{\rm DE}>0$; nevertheless, it can be fulfilled at early cosmological epochs when
\begin{equation}
 a<a_0\, {\rm exp}\left(-\frac{2\Lambda+A}{2A}\right),
\end{equation}
since using equation (\ref{desibling}) we have
\begin{equation}
 \rho_{\rm DE}+3\,p_{\rm DE}=-2\Lambda-A-2A\,{\rm ln}\left(\frac{a}{a_0}\right).
\end{equation}
The QWEC could
be satisfied in the present and future evolution of the universe
only for very small departures with respect a cosmological constant.
Taking into account expressions (\ref{QWEC}) and (\ref{epsilon}), the QWEC would be fulfilled if
\begin{equation}
 A\lesssim 10^{-112}J\times s^{-3}.
\end{equation}
On the other hand, one can consider the non-linear energy conditions. Taking into account
the equation of state (\ref{eqstate}), one can express $p_{\rm DE}$ in terms of $\rho_{\rm DE}$
and simplify the FEC (\ref{FEC}) to
\begin{equation}
 \rho_{\rm DE}\leq-\frac{A}{6}.
\end{equation}
Thus, the FEC was only satisfied when the dark component was sub-dominant, in the early universe.
One can conclude that something similar happens with the DETEC (\ref{DETEC}), since it can be fulfilled
only if $w_{DE}\geq0$. In view of equation (\ref{w}), this is only possible for
\begin{equation}
 \rho_{\rm DE}<0,\qquad {\rm and}\qquad |\rho_{\rm DE}|\leq\frac{A}{3}.
\end{equation}
Moreover, taking again into account (\ref{eqstate}), one can simplify the QFEC (\ref{QFEC}) to
\begin{equation}
  \rho_{\rm DE}\leq\frac{3\epsilon^2}{2A}-\frac{A}{6}.
\end{equation}
Thus, if the right hand side (r.~h.~s.)~of this inequality is positive, the QFEC
associated with the dark fluid would be satisfied for any negative value of $\rho_{\rm DE}$, and for small values of
the energy density. Taking the value given in (\ref{epsilon}) this implies $A<10^{-112}Js^{-3}$.
Moreover, the smaller is the value of $A$, the larger would be the region where the QFEC is
satisfied, being infinitely large for the cosmological constant case $A\rightarrow0$ as it should be expected.

\begin{center}
\begin{table}[h!]
\tbl{In this table we summarise fulfilment/violation of the different energy conditions
 by the dark energy fluid. It must be emphasized that violations of these conditions when this fluid
is subdominant would not generically lead to violations of the conditions for the total cosmic fluid.
The inequalities stands for the regime of fulfilment.}{
{\footnotesize\begin{tabular}{|| l | l | c | c | c | c | c | c | c ||}
\hline
\hline
Model & Epoch 	& NEC    & WEC    & SEC    & QWEC   & FEC & DETEC & QFEC\\
\hline
\hline
\multirow{3}{1cm}{$A>0$}
&  early universe: $\rho_{\rm DE}<0$ 	& \xmark & \xmark & $\rho_{\rm DE}\leq-\frac{A}{2}$ & \xmark & $\rho_{\rm DE}\leq-\frac{A}{6}$ & $\rho_{\rm DE}\leq-\frac{A}{3}$ & \, \cmark $A$ small \\
\cline{2-9}
& present and future & \xmark    & \xmark    & \xmark    & \,\cmark $A$ small   & \xmark & \xmark & \,\cmark $A$ small\\
\cline{2-9}
& close to the singularity & \xmark    & \xmark    & \xmark   & \,\cmark $A$ small   & \xmark & \xmark & \xmark\\
\hline
\hline
\multirow{2}{1cm}{$A<0$}
&cosmic evolution	& \cmark & \cmark & $\rho_{\rm DE}\leq\frac{|A|}{2}$ & \cmark & \cmark & \xmark & \cmark \\
\cline{2-9}
&at the bounce & \cmark    & \cmark    & \cmark     & \cmark    & \xmark & \cmark   & \,\cmark $A$ small \\
\hline
\hline
\end{tabular}}
\label{energycondition}}
\end{table}
\end{center}

\subsection{Dark energy fluid $A<0$}

In this case the dark fluid is always characterised by a positive energy density and the equation of state
fixes
\begin{equation}
 \rho_{\rm DE}+p_{\rm DE}=\frac{|A|}{3}.
\end{equation}
Therefore, the NEC and WEC are clearly satisfied and, therefore, the QWEC.
For the SEC to be satisfied one would need
\begin{equation}
 a\geq a_0 {\rm exp}\left(\frac{2\Lambda-|A|}{2|A|}\right) \equiv a_b\,e^{-1/2},
\end{equation}
where $a_b$ corresponds to the scale factor at the bounce (cf. Sect. II)
Thus, it would generically be satisfied when the dark component dominates the evolution.
Moreover, as in this case $\rho_{DE}\geq0$, then the FEC (\ref{FEC}) can be reduce to the DEC (\ref{DEC}).
Both of them are satisfied for
\begin{equation}
 \rho_{\rm DE}\geq\rho_*= \frac{|A|}{6}.
\end{equation}
That is, these conditions are violated for $\rho<\rho_*$, which implies $a>a_*$ with
\begin{equation}
 a_*=a_b\,e^{-1/6}\sim 0.8\,a_b.
\end{equation}
The violation of these energy conditions close to the bounce could be understood considering that a bounce
may imply a change on the direction of the cosmological arrow of time \cite{Kiefer}. Therefore, a condition which
is based on requiring a causal flux would not be suitably well defined close to this direction flip.
On the other hand, the QFEC would be satisfied in a larger region of the cosmological evolution given by
\begin{equation}\label{qfec3}
 \rho_{\rm DE}\geq\rho_{**}=\rho_*-\frac{3\epsilon^2}{2|A|}.
\end{equation}
Thus, the QFEC associated with the dark fluid would be satisfied
during the whole cosmological evolution if $|A|<3\epsilon\sim 10^{-112}Js^{-3}$, where the r.~h.~s.~of
equation (\ref{qfec3}) is negative,
and would be violated close to the bounce otherwise.
Finally, the DETEC is satisfied in this model only for
\begin{equation}
 \rho_{\rm DE}\leq\frac{|A|}{3},
\end{equation}
thus, it is satisfied in the region close to the bound where this fluid dominates the cosmological dynamics.
In Table \ref{energycondition} we summarise fulfilment/violation of the different energy conditions.
It can be seen that for a fluid with an equation of state with a value of $|A|$ small enough,
that is for a slight departure from the cosmological constant case,
the QFEC is violated only in the far future for the phantom model,
whereas it can be satisfied even at the bounce for the dark
model. This behaviour is reinforcing suggestions of an important role of the QFEC \cite{Martin-Moruno:2013sfa},
as one could think that the singular behaviour of the phantom fluid at infinity is only
possible because the QFEC is violated there. Moreover, one could also argue that fluids
characterized by large positive values of $A$ would not satisfy
the QFEC because the singular behaviour could be felt earlier in the physics of the universe,
as we will see in the next section. On the other hand, as was already pointed out in Ref.~\refcite{Martin-Moruno:2013sfa},
the quantum linear energy conditions don't seem to have the relevance of their non-linear counterparts,
as the QWEC could be fulfilled even when the universe is in a singular state.

\section{Fate of the  bound structures}

In this section, we will study the effects of the accelerated expansion of the universe, as described in Sect. II, on local bound systems. More precisely, we will consider a spherical Newtonian object with mass $M$ and a test particle rotating around it in circular orbit (initially) with a physical radius  $\mathsf{R}$, and assume that both objects are embedded in a spherically symmetric FLRW background. In Refs. \refcite{geodesics,Faraoni:2007es}, the authors have shown that bound systems with a strong enough coupling in a de Sitter background will not comove with the accelerating expansion of the Universe. However it is not the case when general accelerating phases are considered. Therefore, it is necessary to analyse the evolution equations of the physical radius $\mathsf{R}$ of a bound system, or the geodesic equations, at very late-time when the Universe approaches a little sibling of the big rip singularity to find out the outcome of bound systems close to the abrupt event. 
To carry this analysis, we  consider the geodesics of test particles in the appropriate metric that describes the
space-time in the vicinity of a point mass $M$ placed in an expanding background (see for example Ref.~\refcite{geodesics})
\begin{equation}
ds^2 =  -\left(1-\frac{2GM}{\mathsf{R}}\right)dt^2 + a^{2}(t)dr^2 + \mathsf{R}^2d\Omega^2\ ,
\label{metric}
\end{equation}
where $d\Omega^2=d\theta^2+\sin^2\theta d\phi^2$, and $\mathsf{R}(t, r)=ra(t)$. The geodesics corresponding to  metric (\ref{metric}) reads \cite{geodesics}
\begin{align}
\ddot{\mathsf{R}}-\frac{\ddot a}{a}\mathsf{R} + \frac{GM}{\mathsf{R}^2} - \mathsf{R}\dot{\varphi}^2 =0\ ,
\label{EqMotion}
\end{align}
where  $\mathsf{R}^2\dot{\varphi}=L$ with $L$ being the constant angular momentum per unit mass.
We assume that at some initial time $t_1$, the test particle is at circular orbit with radius
$\mathsf{R}_1$, and $\dot{\varphi}(t_1)=\bar{\omega}_0=\sqrt{GM/\mathsf{R}_1^3}$.
Then, by introducing the dimensionless parameters
\begin{equation}\
\omega_0 := \bar{\omega}_0 t_1=\sqrt{\frac{GMt_1^2}{\mathsf{R}_1^3}}\ ,\quad\quad   R:=\frac{\mathsf{R}}{\mathsf{R}_1}\ , \quad\quad   \tau:=\frac{t}{t_1} \ ,
\end{equation}
the radial equation of motion (\ref{EqMotion}) for a test particle in the Newtonian limit is given in a dimensionless form as
\begin{equation}
R^{\prime\prime}\  =\  \frac{a^{\prime\prime}}{a}R+ \frac{\omega_0^2}{R^3} - \frac{\omega_0^2}{R^2}  \ .
\label{EqMotion2}
\end{equation}
where a prime stands for a derivative with respect to $\tau$.

We are interested on the fate of bounded structure close to the little sibling of the big rip. Therefore, the expansion of the universe is described by the Friedmann equation (\ref{scalefactor2}) with solution
(\ref{Hubbleinfinity}). By plugging this solution into the  Raychaudhuri equation we obtain
\begin{align}
\frac{a^{\prime\prime}}{a}\  \approx\ &    \left(\frac{\Omega_x}{2}H_0(t-t_1) + \sqrt{\Omega_\Lambda+\Omega_xx_1}\right)^2 H_0^2t_1^2 + \frac{\Omega_x}{2}H_0^2 t_1^2\ =:\   \lambda^{2}(t)\  .
\label{Raychadhuri2}
\end{align}
Then, by substituting Eq. (\ref{Raychadhuri2}) in Eq. (\ref{EqMotion2}), we obtain the equation of motion of a two body gravitating system in our expanding universe as
\begin{align}
R^{\prime\prime} + \frac{\omega_0^2}{R^2}\left(1-\frac{1}{R}\right) -  \lambda^{2}(t) R = 0.
\label{EqMotion3-b}
\end{align}
The evolution of the radius of the bound system can be evaluated explicitly by numerically solving the equation of motion (\ref{EqMotion3-b}) (see for example Ref. \refcite{geodesics}).
An alternative way to analyse the stability of the system is by calculating the effective potential minimum of the above system and by studying the evolution of the potential minimum;.i.e. does the minimum last for ever or does it disappear? If the minimum last for ever, the gravitational structure will be stable, i.e. it will remain bound in the future. However, if  the minimum of the potential disappears as the universe expands then the bound system will disappear \cite{geodesics}. We will follow precisely this approach to study the  dynamics of the bound system (\ref{EqMotion3-b}).

Using the Newtonian form of the equation of motion $R^{\prime\prime}=f(R)$ for the gravitational force $f(R)$, we can
derive the effective potential that determines the dynamics of the bound system (\ref{EqMotion2}) as  \cite{geodesics}
\begin{align}
V_{\rm eff}\ =\  -\frac{\omega_0^2}{R} + \frac{\omega_0^2}{2R^2}-\frac{1}{2}\lambda^2(t)R^2\ .
\label{potential}
\end{align}
where
\begin{align}
\frac12 \dot{R}^2\ =\ - V_{\rm eff}
\label{EqMotionpot}
\end{align}

\begin{figure}
\begin{center}
\includegraphics[height=2.7in]{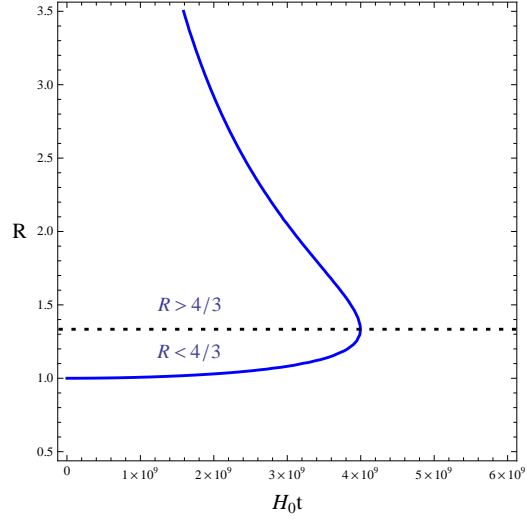}
\caption{{\footnotesize Numerical values of the  radius $R_{\rm ext}$, given by Eq.~(\ref{exteremum}),  for the Solar system (where $\omega_0 \sim 3.5\times 10^{6}$).
The region $R<4/3$ on the curve shows the solutions for which the effective potential of the bound system is minimum.
For simplicity, we have assumed that  $x_1=-\ln(1+z_m)$, $\Omega_m= 0.315$, $\Omega_x=10^{-3}$ and $t_1\simeq7.97$ Gyrs, where $z_m$ and $t_1$ corresponds to the redshift when the universe starts accelerating.}}
\end{center}
\label{F2}
\end{figure}

The extremum  of the effective potential (\ref{potential}) can be obtained by solving the equation\footnote{\label{footnoteIV} The discriminant of the quartic potential on the r.~h.~s.~ of equation (\ref{exteremum}) reads $D=-27Q^4+256Q^6$ \cite{Abramowitz}. There are real solutions only when $D<0$. We will stick to this case. \label{footnoteIV}} \cite{geodesics}:
\begin{align}
R_{\rm ext} -1\  =\  Q^2(t) R_{\rm ext}^4\ ,
\label{exteremum}
\end{align}
where $Q(t):=\lambda(t)/\omega_0$, and $R_{\rm ext}(t)$ is the time  dependent radius of the bound system.
Fig.~\ref{F2} shows values of the radius $R_{\rm ext}$, given by Eq.~(\ref{exteremum}), for different values of time.

\begin{figure}
\begin{center}
\includegraphics[height=2.1in]{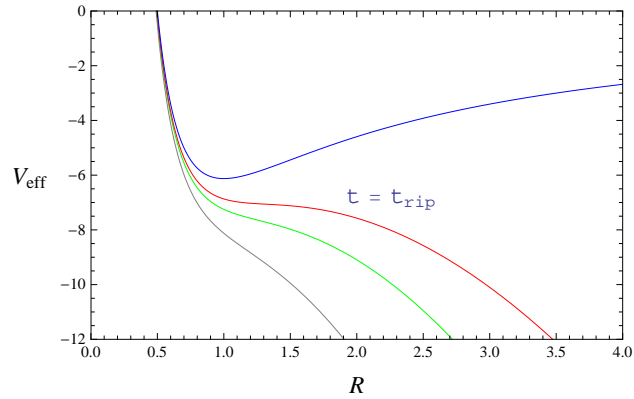}
\caption{{\footnotesize A pictorial behaviour of the effective potential $V_{\rm eff}(R)$  when approaching the time $t_{\rm rip}$ as can be seen at that point the minimum of the potential disappears and the  bound system
as well.}}
\end{center}
\label{Potential1}
\end{figure}

Using the stability condition  $d^2V_{\rm eff}/dR^2(R_{\rm min})>0$ together with Eq.~(\ref{exteremum}), we find that the minimum of the potential is located
in the range of the radius $R_{\rm min} < 4/3$; similarly, the condition $d^2V_{\rm eff}/dR^2(R_{\rm max})<0$  implies that the maximum of the potential occurs in the range $R_{\rm max} > 4/3$.
As we have mentioned earlier, we assume that initially  the effective potential reaches some minimum value at $t_1$; i.e. the system is initially on a circular orbit $R_{\rm min}(t_1)$.
We take $t_1$ as the time when the universe starts accelerating by using the method given in Ref.~\refcite{geodesics}.
In Fig.~\ref{Potential1} we show an example of the evolution of the effective potential $V_{\rm eff}$ with respect to the physical radius $R$.

We next estimate the time when the structure ceases to be bound. On the one hand, Eq. (\ref{exteremum}) has a solution only when $Q^{2}(t)\leq (27/256) =: Q_c^2$ (see footnote \ref{footnoteIV}).
On the other hand, the system becomes unbound when the minimum of the potential (\ref{potential}) disappears; the time $t_{\rm rip}$ when the system becomes unbound is given by
$Q(t_{\rm rip})=Q_c$. Therefore, taking into account Eq. (\ref{Raychadhuri2}), we obtain
\begin{equation}
  t_{\rm rip}\  \approx \ t_1 + \frac{2}{\Omega_xH_0}\left[\sqrt{\frac{Q_c^2\omega_0^2}{t_1^2H_0^2}-\frac{\Omega_x}{2}} -\sqrt{\Omega_\Lambda+\Omega_xx_1}\right].
\label{t-rip}
\end{equation}

We will next compare the dissociation time for bound structure in our model and a model with a big rip singularity.
For this goal, we next review the results for the big rip model \cite{Caldwell:2003vq,geodesics}.
We consider a FLRW universe filled with a phantom energy component with a constant equation of state $\tilde{w}$.
The corresponding Friedmann equation
in the dark energy dominated era can be approximated by
\begin{equation}
\frac{\dot{a}^2}{a^2}\  \approx\  H_0^2(1-\Omega_m)a^{-3(1+\tilde{w})}.   \label{friedmannfuture2}
\end{equation}
By integrating this equation,
we find the scale factor as
\begin{align}
a(t) \ = \  \left[1- \frac{H_0}{\alpha} \sqrt{1-\Omega_m}(t-t_1)\right]^{-\alpha}\  ,
\label{scalefactor-BR}
\end{align}
where $\alpha:=\frac{2}{3|1+\tilde{w}|}$ and we have considered $a(t_1)=1$.
This equation implies that, at $t_{\rm BR}$ where
\begin{equation}
t_{\rm BR}-t_1\  \approx \ \frac{\alpha}{H_0\sqrt{1-\Omega_m}} \ ,
\label{t-BRSing}
\end{equation}
a big rip singularity happens in the future   \cite{Caldwell:2003vq}.
Then, by substituting the solution (\ref{scalefactor-BR}) into the Raychaudhuri equation,  the corresponding $\tilde{\lambda}^2(t):=a^{\prime\prime}/a$ for this model reads
\begin{align}
\tilde{\lambda}^2(t)\ =\   H_0^2t_1^2\left(1-\Omega_m\right)\left(1+\frac{1}{\alpha}\right) \left[1- \frac{H_0}{\alpha}\sqrt{1-\Omega_m}(t-t_1)\right]^{-2}\ .
\label{lambda-BR}
\end{align}
The radial motion of a bound system in a universe with a future big rip singularity is given by Eq.~(\ref{EqMotion3-b}) after substituting $\lambda(t)$ by $\tilde{\lambda}(t)$.
Following a similar procedure  to the one previously used, the time $\tilde{t}_{\rm rip}$ for the dissociation of a bound system can be obtained by setting $\tilde{Q}^2(\tilde{t}_{\rm rip}):=\tilde{\lambda}^2(\tilde{t}_{\rm rip})/\omega_0^2=Q_c^2$.
Consequently, we obtain
\begin{align}
\tilde{t}_{\rm rip} - t_1\ =\  \ \frac{\alpha}{H_0\sqrt{1-\Omega_m}} -\frac{t_1\sqrt{\alpha(\alpha+1)}}{\omega_0Q_c}\ .
\label{t-rip-BR}
\end{align}
By combining Eqs. (\ref{t-BRSing}) and (\ref{t-rip-BR})  we can rewrite $\tilde{t}_{\rm rip}$ as
\begin{align}
\tilde{t}_{\rm rip}\ =\  t_{\rm BR} -\frac{t_1\sqrt{\alpha(\alpha+1)}}{\omega_0Q_c}\ ,
\label{t-rip-BR2}
\end{align}
which indicates that the dissociation of a bound system occurs at a timescale $t_1\sqrt{\alpha(\alpha+1)}/(\omega_0Q_c)$  earlier than the big rip.
\begin{center}
\begin{table} [h!]
\tbl{The dissociation times differences $\tilde{t}_{\rm rip}-t_{1}$ (in the model with big rip; cf. Eq.~(\ref{t-rip-BR})) and $t_{\rm rip}-t_{1}$ (in our herein model; cf.  Eq.~(\ref{t-rip})), for three bound systems (with $H_0=70$ km.sec$^{-1}$.Mpc$^{-1}$, and $\tilde{w}=-1.01$ for phantom energy in the big rip model).
We took ~$t_1\simeq7.97$ Gyrs, as the time when the universe starts accelerating.
We further used $\omega_0=3.5\times 10^{6}$ for Solar system, $\omega_0=182$ for Milky Way galaxy and $\omega_0=4.15$ for Coma cluster \cite{geodesics}.}{
\begin{tabular} { |c |c |c |c || }
\hline  \hline
~~System ~~& ~~ $\tilde{t}_{\rm rip}-t_{1}$~(yrs) ~~&  ~~ $t_{\rm rip}-t_{1}$~(yrs) ~~  \\
\hline\hline
Solar system &  $1.13 \times 10^{12}$  & $5.75\times 10^{19}$  \\

Milky Way galaxy & $1.12 \times 10^{12}$ & $2.88 \times 10^{15}$  \\

Coma Cluster &  $7.30 \times 10^{11}$  &  $4.32 \times 10^{13}$ \\
\hline \hline
\end{tabular}}
\label{fixed-points}
\label{dissociationtime}
\end{table}
\end{center}
We compare  the dissociation time of a bound structure in both models in Table \ref{dissociationtime}.
As could be expected the milder character of the little sibling of the big rip as compared with the big rip, leaves also its imprints on the dissociation time of a bound structure;
i.e. it would take a longer time for a given structure to be dissociated in our universe if the doomsday would correspond to the little sibling of the big rip rather than the big rip.

Before concluding this section, we would like to stress that the analysis we have carried out and which is based on the evolution 
equation (\ref{EqMotion2}) is valid within a Newtonian approximation and under a weak field limit as clearly stated in Ref. \refcite{geodesics}. On the other hand, there are 
many choices of metrics one can use to interpolate between a Schwarzschild and a FLRW metric (cf.  Ref.  \refcite{Faraoni:2007es}).  
The results we have obtained may be improved if a better interpolating metric is chosen. However, we expect the approximation in equation (\ref{metric})  to be still valid as the universe 
approaches the little sibling of the big rip because the matter part (the terms proportional to $GM$ in the evolution equation of the 
bound structure) is small as 
compared with the expansion terms within such conditions. 
Therefore, the use of the evolution equation (\ref{EqMotion}) is quite fair in our analysis.

\section{Conclusions}

We present a dark energy model which deviates slightly from the $\Lambda $%
CDM at present, cf. Eq.~(\ref{eqstate}). Its energy density has a
logarithmic correction of the scale factor as compared with a pure
cosmological constant, such that its influence on the past of the universe
is negligible in comparison with the other matter components of the
universe. At present it behaves like a cosmological constant. However, its
future behaviour is drastically different from that found on a $\Lambda $%
CDM. In fact, the universe would be no longer asymptotically de Sitter as in the $\Lambda$CDM scenario but would face
what we named the little sibling of the big rip if $A$ is positive (see Eq.~(%
\ref{eqstate})). At this event the Hubble rate and the scale factor blow up
but the cosmic derivative of the Hubble rate does not. This abrupt event
takes place at an infinite cosmic time where the scalar curvature explodes.
We use the nomenclature of the little sibling of the big rip because (i) the
Hubble rate diverges like in the big rip singularity and (ii) it takes the
universe an infinite cosmic time to reach the event like what happens with
the little rip event. We have as well analysed the outcome of bound structure in the universe if its doomsday would corresponds
to a little sibling of the big rip. It turns out that even though the event seems to be harmless as it takes place in the infinite future, the bound structures in the universe would be unavoidably destroyed on a finite cosmic time from now (cf. Table~\ref{dissociationtime}).

In addition, we expect that a possible way to smooth the abrupt event corresponding to the
little sibling of the big rip singularity could be through the  inclusion of ultra-violet and
infra-red modification of general relativity as the event happens in the future and
at very high energies \cite{Ouali:2011, Ouali:2012, BouhmadiLopez:2013}.

In order to complete our analysis, we have also shown that a universe with a negative value of $A$ bounces in its future.
The smaller is the dimensionless energy density $\Omega _{x}$,
quantifying the deviation of the dark energy model (\ref{eqstate}) from a cosmological constant, the latter
would the
bounce take place.

We have as well given a clear motivation for considering this model
from the point of view of the energy conditions, since the resulting cosmological model violates
the WEC only in a controlled way even in the phantom dominated future expansion,
therefore, satisfying the QWEC.
A deeper consideration of this fact could lead us to doubt about the
utility of the QWEC (see also Ref. \refcite{Martin-Moruno:2013wfa}), as we have proven that
this condition can be satisfied even when a singular event takes place.
On the other hand, the QFEC is violated only
close to the singularity for small enough values of $A$
(that is, for large enough $t_{\rm rip}$), where the universe
suffers the singular effects. This interesting behaviour could be suggesting that a more fundamental
theory able to resolve singularities could be compatible with (or it may even imply) the QFEC. Moreover,
the fact that this conditions can be satisfied at the turnaround of the model with $A<0$ for small values of
$|A|$,
it is only confirming the weaker character of this kind of events, which are not expected to be smoothed
by quantum effects.

\section*{Acknowledgments}

The work of M.B.L. was supported by the Basque Foundation for Science IKERBASQUE and the Portuguese Agency ``Funda\c{c}\~{a}o para a Ci\^{e}ncia e Tecnologia" (FCT) through an
Investigador FCT Research contract, with reference IF/01442/2013/CP1196/CT0001. A.E. and T.O. are supported by CNRST, through the fellowship URAC 07/214410.  P.M.M. is supported by FCT. Y.T. is supported by a PDJ fellowship from the Brazilian agency CNPq. This work was supported by FCT (Portugal) through the projects PTDC/FIS/111032/2009,   PEst-OE/MAT/UI0212/2014 and EXPL/FIS-AST/1608/2013 and partially by the Basque government through the Grant No. IT592-13.


\end{document}